\documentstyle[graphicx]{mn}	

\newif\ifAMStwofonts

\def\refit{}\def\refbf{}
\def\myref#1,#2,#3.{{\refit #1\/}, {\refbf #2}, #3\par\noindent}
\def\aa,#1,#2.{\myref{A\&A},#1,#2.}
\def\acta,#1,#2.{\myref{Acta Astron.},#1,#2.}
\def\annrev,#1,#2.{\myref{ARAA},#1,#2.}
\def\araa,#1,#2.{\annrev,#1,#2.}
\def\aj,#1,#2.{\myref{AJ},#1,#2.}
\def\apj,#1,#2.{\myref{ApJ},#1,#2.}
\def\apjsupp,#1,#2.{\myref{ApJS},#1,#2.}
\def\apspsci,#1,#2.{\myref{Ap\&SS},#1,#2.}
\def\aasupp,#1,#2.{\myref{AA\ Supp.},#1,#2.}
\def\ica,#1,#2.{\myref{Icarus},#1,#2.}
\def\grg,#1,#2.{\myref{GRG},#1,#2.}
\def\jaa,#1,#2.{\myref{J.\ Astr.\ Astrophys.},#1,#2.}
\def\mnras,#1,#2.{\myref{MNRAS},#1,#2.}
\def\nat,#1,#2.{\myref{Nature},#1,#2.}
\def\newa,#1,#2.{\myref{NewA},#1,#2.}
\def\pasp,#1,#2.{\myref{PASP},#1,#2.}
\def\pasj,#1,#2.{\myref{PASJ},#1,#2.}
\def\physrev,#1,#2.{\myref{Phys.\ Rev.},#1,#2.}
\def\physrevlett,#1,#2.{\myref{Phys.\ Rev.\ Lett.},#1,#2.}
\def\physrevD,#1,#2.{\myref{Phys.\ Rev.\ D},#1,#2.}
\def\procroy,#1,#2.{\myref{Proc.\ Roy.\ Soc.},#1,#2.}
\def\revmod,#1,#2.{\myref{Rev.\ Mod.\ Phys.},#1,#2.}
\def\sova,#1,#2.{\myref{SvA},#1,#2.}

\def\etal{{\it et\thinspace al\/}}

\def\ie{{i.e}}
\def\eg{{e.g}}

\def\frac(#1/#2){{\textstyle#1\over\textstyle#2}}

\edef\vbar{|}
\def\subrm#1{_{\rm #1}}
\catcode`|=\active \let|=\subrm

\def\mod#1{{\left\vbar#1\right\vbar}}
 
\def\pd#1#2{{\upartial#1\over \upartial#2}} \def\lr#1{\left(#1\right)}
\def\td#1{{\rm d}#1}
\def\d#1#2{{\td#1\over\td#2}}
\def\dts#1#2{{\textstyle\td#1\over\textstyle\td#2}}

\newcount\psw 
\def\computelabs{
  \psw=\number\epsfxsize \divide\psw by 65536 
}%
\newbox\labelbox
\newdimen\yy\newdimen\xx
\newdimen\hair\hair=3pt

\def\xlabel#1#2#3{%
\setbox\labelbox\hbox{$#1$}%
\getkerns{#2}{#3}%
\advance\xx by -.5\wd\labelbox%
\putlabel{0}%
}

\def\ylabel#1#2#3{%
\setbox\labelbox\hbox{$#1$}%
\getkerns{#2}{#3}%
\advance\yy by .5\wd\labelbox%
\putlabel{90}%
}

\def\putlabel#1{%
\vbox to 0pt{\kern-\yy\hbox to 0pt{\kern\xx%
\rotatebox{#1}{\box\labelbox}%
\hss}\vss}%
\ifvmode\nointerlineskip\fi%
}

\def\getkerns#1#2{%
\yy=#2 pt 
\divide \yy by \mag%
\multiply \yy by 1010%
\multiply \yy by \the\psw%
\xx=#1 pt 
\divide \xx by \mag%
\multiply \xx by 1010%
\multiply \xx by \psw%
\advance \xx by .5\hsize
\advance \xx by -.5\epsfxsize
}

\def\reference{\item}

\def\Msun{{\ifmmode M_{\sun} \else $M_{\sun}$ \fi}}
\def\msun{{\ifmmode m_{\sun} \else $m_{\sun}$ \fi}}
\def\Lsun{{\ifmmode L_{\sun} \else $L_{\sun}$ \fi}}
\def\rsun{{\ifmmode r_{\sun} \else $r_{\sun}$ \fi}}
\def\Rsun{{\ifmmode R_{\sun} \else $R_{\sun}$ \fi}}
\def\kms{\hbox{km\,s$^{-1}$}}
\def\pc{\hbox{pc}}

\def\y{\hbox{y}}
\def\magsec{\hbox{mag/arcsec$^2$}}

\def\ndot{{\dot{N}}}
\def\ndotl{{\dot{N}_{<}}}
\def\ndotg{{\dot{N}_{>}}}
\def\M{{M}}
\def\Lstar{{L_*}}
\def\thlc{{\theta|{lc}}}
\def\rcrit{{r|{crit}}}
\def\raa{{r|{crit,1}}}
\def\r2c{{r|{crit,2}}}
\def\rg{{\rm{RG}}}
\def\tr{{t}}

\def\FR{FR}
\def\CK{CK}
\def\FRone{Frank \& Rees (1976) (\FR)}
\def\CKone{Cohn \& Kulsrud (1978) (\CK)}

\input epsf

\ifoldfss
  \ifCUPmtlplainloaded \else
    \NewTextAlphabet{textbfit} {cmbxti10} {}
    \NewTextAlphabet{textbfss} {cmssbx10} {}
    \NewMathAlphabet{mathbfit} {cmbxti10} {} 
    \NewMathAlphabet{mathbfss} {cmssbx10} {} 
  \fi
  \ifAMStwofonts
    \ifCUPmtlplainloaded \else
      \NewSymbolFont{upmath} {eurm10}
      \NewSymbolFont{AMSa} {msam10}
      \NewMathSymbol{\upi}     {0}{upmath}{19}
      \NewMathSymbol{\umu}     {0}{upmath}{16}
      \NewMathSymbol{\upartial}{0}{upmath}{40}
      \NewMathSymbol{\leqslant}{3}{AMSa}{36}
      \NewMathSymbol{\geqslant}{3}{AMSa}{3E}

    \fi
  \fi
\fi 

\ifnfssone
  \newmathalphabet{\mathit}
  \addtoversion{normal}{\mathit}{cmr}{m}{it}
  \addtoversion{bold}{\mathit}{cmr}{bx}{it}
  \newmathalphabet{\mathbfit} 
  \addtoversion{normal}{\mathbfit}{cmr}{bx}{it}
  \addtoversion{bold}{\mathbfit}{cmr}{bx}{it}
  \newmathalphabet{\mathbfss} 
  \addtoversion{normal}{\mathbfss}{cmss}{bx}{n}
  \addtoversion{bold}{\mathbfss}{cmss}{bx}{n}
  \ifAMStwofonts
    \ifCUPmtlplainloaded \else
      \UseAMStwoboldmath
      \makeatletter
      \new@mathgroup\upmath@group
      \define@mathgroup\mv@normal\upmath@group{eur}{m}{n}
      \define@mathgroup\mv@bold\upmath@group{eur}{b}{n}
      \edef\UPM{\hexnumber\upmath@group}
      \new@mathgroup\amsa@group
      \define@mathgroup\mv@normal\amsa@group{msa}{m}{n}
      \define@mathgroup\mv@bold\amsa@group{msa}{m}{n}
      \edef\AMSa{\hexnumber\amsa@group}
      \makeatother
      \mathchardef\upi="0\UPM19
      \mathchardef\umu="0\UPM16
      \mathchardef\upartial="0\UPM40
      \mathchardef\leqslant="3\AMSa36
      \mathchardef\geqslant="3\AMSa3E
    \fi
  \fi
\fi 

\ifnfsstwo
  \DeclareMathAlphabet{\mathbfit}{OT1}{cmr}{bx}{it}
  \SetMathAlphabet\mathbfit{bold}{OT1}{cmr}{bx}{it}
  \DeclareMathAlphabet{\mathbfss}{OT1}{cmss}{bx}{n}
  \SetMathAlphabet\mathbfss{bold}{OT1}{cmss}{bx}{n}
  \ifAMStwofonts
    \ifCUPmtlplainloaded \else
      \DeclareSymbolFont{UPM}{U}{eur}{m}{n}
      \SetSymbolFont{UPM}{bold}{U}{eur}{b}{n}
      \DeclareSymbolFont{AMSa}{U}{msa}{m}{n}
      \DeclareMathSymbol{\upi}{0}{UPM}{"19}
      \DeclareMathSymbol{\umu}{0}{UPM}{"16}
      \DeclareMathSymbol{\upartial}{0}{UPM}{"40}
      \DeclareMathSymbol{\leqslant}{3}{AMSa}{"36}
      \DeclareMathSymbol{\geqslant}{3}{AMSa}{"3E}
    \fi
  \fi
\fi 

\ifCUPmtlplainloaded \else
  \ifAMStwofonts \else 
    \def\upi{\pi}
    \def\umu{\mu}
    \def\upartial{\partial}
  \fi
\fi

\title[] 
{Tidal disruption rates of stars in observed galaxies}
\author[]
{D. Syer and A. Ulmer
\thanks{E-mail: (syer, aulmer)@mpa-garching.mpg.de} \\
	Max-Planck-Institut f\"ur Astrophysik
	Karl-Schwarzschild-Strasse 1, 85748 Garching, Germany}
\date{Accepted ........
      Received .......;
      in original form .......}

\pagerange{\pageref{firstpage}--\pageref{lastpage}}
\pubyear{1998}

\begin{document}
\maketitle
\label{firstpage}

\begin{abstract}
We derive the rates of capture $\ndot$ of main sequence turn off stars
by the central massive black hole in a sample of galaxies from
Magorrian \etal{.} 1998. The disruption rates are smaller than
previously believed with $\ndot \sim 10^{-4}-10^{-7}$ per galaxy.  A
correlation between $\ndot$ and black hole mass $\M$ is exploited to
estimate the rate of tidal disruptions in the local universe.
Assuming that all or most galaxies have massive black holes in their
nuclei, this rate should be dominated by sub-$\Lstar$ galaxies.  The
rate of tidal disruptions could be high enough to be detected in
supernova (or similar) monitoring campaigns---we estimate the rate of
tidal disruptions to be $0.01-0.1$ times the supernova rate.  We have
also estimated the rates of disruption of red giants, which may be
significant ($\ndot \ga 10^{-4}\y^{-1}$ per galaxy) for $\M\ga
10^8\Msun$, but are likely to be harder to observe---only of order
$10^{-4}$ times the supernova rate in the local universe.  In
calculating capture rates, we advise caution when applying scaling
formulae by other authors, which are not applicable in the physical
regime spanned by the galaxies considered here.
\end{abstract}


\section{Introduction}
A number of early type galaxies and spiral bulges are now thought to
contain massive black holes in their nuclei (Magorrian \etal{.}
1998).  Direct evidence is also now available which supports the idea
that active galaxies are powered by massive black holes (Reynolds \&
Fabian 1997, Fabian \etal{.} 1998).  It can be argued that many if not
all galaxies could be expected to have undergone an active phase and
to possess a massive black hole (Haehnelt \& Rees 1993).  Magorrian
\etal{.} measure black hole masses in the range $\M=10^6-10^9\Msun$,
and report that $\M$ is correlated with the mass $\cal M$ of the
underlying hot stellar system, with $\M/{\cal M} = x \approx 0.006$.

A main sequence star of roughly solar type can be tidally disrupted by
a black hole with mass $\M\la2 \times 10^8\Msun$ (Hills 1975).
Larger black holes swallow main
sequence stars whole, but red giants are susceptible to tidal
disruption because of their lower density.  \FRone{}, Young, Shields
\& Wheeler (1977) considered a system composed of a black hole and an
isothermal sphere of stars and derived analytic expressions for the
rate of capture of stars.  \CKone{} were able to calibrate \FR{} using
a more detailed numerical calculation.  Estimates of capture rates in
the literature are often based on equation (66) of \CK{} (\eg{.}
Cannizzo, Lee \& Goodman 1990, Kormendy \& Richstone 1995).  We 
argue that such estimates are often wrong because the physical
conditions in galactic nuclei are different from those anticipated by
\CK{} (who were in any case more concerned with globular clusters).

The next section summarises briefly what is known about the
observational signatures of tidal disruption.  Section \ref{conesec}
establishes notation by reviewing the analytic theory of \FR{} and
sets out our method for calculating capture rates in the Nuker
galaxies.  We calibrate our calculation against the result of \CK{}.
In Section \ref{datsec} we describe the data and list the quantities
we derive from them, including capture rates.  We discuss our results
in Section \ref{discuss}.

\section{Observability}
The observable consequences of such an event have been discussed by
Rees (1988), Evans \& Kochanek (1989), Canizzo, Lee \& Goodman (1990)
and Ulmer (1997). The general expectation for the disruption of a main
sequence star is a flare with bolometric luminosity near the Eddington
limit. In comparison, the luminosity from the central $r<1\,$arcsec of
all the Nuker galaxies is small compared to the Eddington luminosity
of the black hole ($\la10^{-4}L|{Edd}$, Table \ref{dattab}).  The
bolometric luminosity should be close to the Eddington limit because
of the predicted mass accretion rates are above the Eddington
accretion rate for black holes with mass up to about $3\times
10^7~\Msun$. Above that the predicted mass accretion rate divided by
the Eddington mass accretion rate falls as
$M^{3/2}$ (Ulmer~1997).  There is much uncertainty as to how optically
bright a tidal disruption event will be, in optical, but the minimum
luminosity expected for typical disruptions of main sequence stars is
$10^{-3}{-}10^{-2} L|{Edd}$ for the U and V bands (Ulmer
1997). Typical durations of the brightest phase will be
$0.1{-}2$~years depending on the black hole mass. Main sequence tidal
disruption events should therefore be relatively easy to detect as a
byproduct of supernova searches (Section \ref{detsec}) or other
searches which observe the same galaxies many times over (\eg{.}
Southern strip of the SDSS).

Red giant disruption events are generally much longer lasting, and
therefore fainter, than main sequence events. Red giants typically are
disrupted by growing onto the loss cone (Section \ref{rgsec}), so at
disruption, they have pericenter equal to the tidal disruption radius.
Consequently, an average red giant (with $R\sim 100 \Rsun$)
disruption has a approximate duration of $100 M_6^{1/2}$~years and a
produces a peak mass accretion rate no larger than $\sim
10^{-3}~M_6^{-1/2} \Msun$~yr$^{-1}$ (see Ulmer~1997, eqs 3,5). The
Eddington accretion rate, which would supply enough mass to provide
the Eddington luminosity, is $3 \times 10^{-2} M_6
\epsilon^{-1}_{.1}$, where $0.1\times\epsilon_{.1}$ is the efficiency of
rest mass to light conversion in the accretion process. This means
that red giant disruption could keep a small, $10^6~\Msun$ black
hole illuminated at (1/30th) its Eddington rate for up to a thousand
years.

Just how bright a red giant disruption would appear in optical
bands is difficult to determine, but a simple estimate is as follows.
If most of the energy is released from within a few Schwarzschild
radii for both the main sequence and red giant disruptions, and the
bolometric luminosity of the red giant disruptions is 1/30 that of
the main sequence disruptions for a $10^6~\Msun$~black hole, then
\begin{equation}
{1\over 30} \sim {L_\rg \over L|{MS}} \sim { T|{eff,\rg}^4
\over T|{eff,MS}^4}.
\end{equation}
Therefore, the ratio of temperatures is $(1/30)^{1/4}\sim
0.4$. Because the optical bands lie in the Rayleigh-Jeans tail of the
spectrum where the luminosity scales linearly with $T$, we would at
first guess expect that red giant disruptions would be nearly (\ie{.}
$\sim 0.4$ times) as bright as main sequence events.  Around more
massive black holes with $M\sim 10^8\Msun$, red giants may be
significantly less luminous if the accretion becomes advection
dominated as appears to be the case for many extra-galactic black
holes with accretion rates far below the Eddington accretion rate
(\eg{.} NGC~4258 (Lasota~\etal{.} 1996), M87 (Reynolds \etal{.}
1996)).  Around very massive black holes with $M\sim 10^{10} M_\odot$,
the timescale for return of the disruption material to the black hole
becomes comparable with the interval between successive tidal
disruptions (see Figure~\ref{ndotm}). Red giant disruptions would then
provide a small, nearly constant flow of material onto the black hole
of $\sim 10^{-3} M_\odot {\rm yr}^{-1}$ corresponding to a luminosity
of $\sim 10^{-5} \epsilon_{.1} L|{Edd}$

\section{Theory of capture rates}\label{conesec}
Consider a spherical cluster of stars of mass $m_*$ with density
$\rho(r)$, and isotropic velocity dispersion $\sigma(r)$.  Typically
$\rho(r)$ is in the form of two power laws, separated by a break
radius $r|b$.  The velocity dispersion is then given by Jeans'
equations:
\begin{equation}
\d{\rho\sigma^2}{r} + {G \rho {\cal M}(r)\over r^2} = 0
\label{jeans}
\end{equation}
where ${\cal M}(r)$ is the mass enclosed within radius $r$, including the
black hole, and $G$ is Newton's constant.
The black hole exerts
an influence on the stars inside a
sphere of influence with radius
$r|a$ given by
\begin{equation}
{G\M\over r} - \sigma(r)^2 = 0,
\label{infl}
\end{equation}
where $\M$ is the black hole mass.
Within the sphere of influence ($r<r|a$)
\begin{equation}
\sigma^2\approx {G\M\over (1+p) r},
\label{sigra}
\end{equation}
where the logarithmic slope of the density is $-p$ (\ie{.} $\rho\sim
r^{-p}$, with $p$ possibly dependent on radius).  An important derived
quantity is the two-body relaxation timescale, given by
\begin{equation}
t|r = {\sigma^3\over \Lambda G^2\rho m_*},
\label{tr}
\end{equation}
where $m_*$ is the typical stellar mass and
$\Lambda$ includes dimensionless factors of order unity as well
as the Coulomb logarithm.  We use the numerical value of $t|r$ given by
Binney \& Tremaine 1987, equation (8-71).
Since $t|r$ is not
very sensitive to $\Lambda$ we do not attempt to include the
dependence of $\Lambda$ on $r$.  Instead we take 
\begin{eqnarray}
\lefteqn{t|r =}\nonumber\\
&&\kern-1em{1.8\times10^{12} \y\over \ln(0.4N) }
\lr{\sigma\over100\kms}^3
\lr{\Msun\over m_*} \lr{10^4\Msun\pc^{-3}\over\rho},
\label{trnum}
\end{eqnarray}
where $N$ is the number of stars within the characteristic radius
$r|b$.

\subsection{The loss cone}
Suppose that stars are removed from the system if they come within a
distance $q$ of the centre of the cluster.  If there is a massive
black hole at the centre, $q$ will be the larger of the tidal radius
$r|T \approx R_\star (\M /M_\star)^{1/3} $ (Hill 1975) or the
Schwarzschild radius of the black hole.  For stars at the main
sequence turn off, $q=r|T$ when $\M\la2 \times 10^8\Msun$.  A star at
radius $r$ will be removed if its angular momentum is small enough.
Such stars are said to populate a `loss-cone' (Lightman \& Shapiro
1977, Young, Shields \& Wheeler 1977, \FR), the angular size of which
is given at radius $r$ by
\begin{equation}
\theta|{lc}^2 = {q\over r^2} {G\M\over\sigma^2}
\label{thlc}
\end{equation}
(provided $\sigma^2 \ll \M/q$). Even at the edge of the sphere of
influence, the loss cone is quite small with
$\theta|{lc} \sim 10^{-7}-10^{-5}$.

The loss process is usually divided into two regimes according to the
angle $\theta|d$ through which a star is scattered in a dynamical time
\begin{equation}
t|d = r/\sigma.
\label{td}
\end{equation}
In the `diffusive' regime
$\theta|d<\thlc$, and on timescales much longer than $t|d$ the loss
cone is empty, because a star which finds itself in the loss cone is
removed within a dynamical time.  In this case, the capture rate is
limited by diffusion into the loss cone, and is given, per star, by
\begin{equation}
\d{\ndotl}{N} = {1\over \ln(2/\thlc) t|r}
\label{diffuse}
\end{equation}
(Lightman \& Shapiro 1977),
where $N(r)$ is the number of stars contained within a radius $r$.

In the other regime, $\theta|d>\thlc$, the loss cone is always full.
Since we assume isotropic velocity dispersions, the fraction of stars
in the loss cone at any time is just $\thlc^2$.  Thus in this case the
capture rate per star is
\begin{equation}
\d{\ndotg}{N} = {\thlc^2 \over t|d}.
\label{full}
\end{equation}
The rates (\ref{diffuse}) and (\ref{full}) are equal at a radius
$r=\rcrit$, where
\begin{equation}
{G\rho r^3\over \sigma^2} = q {\M\over m_*} \Lambda|{lc}
\label{rcrit}
\end{equation}
where we have written $\Lambda|{lc}=\ln(2/\thlc)/\Lambda$.  The radial
dependence of $\Lambda|{lc}$ is weak, so we set it to a constant equal
to its value at $r|b$.

The first step in calculating the capture rate is to solve equation
(\ref{rcrit}) to find $\rcrit$.  Then we can find the total loss rate,
$\ndot$, by integrating equations (\ref{diffuse}) and (\ref{full}) over the
cluster:
\begin{equation}
\ndot = \ndotl \;+\; \ndotg,
\label{ndot}
\end{equation}
with
\begin{equation}
\ndotl = 4\pi \int\limits_0^\rcrit\; {\Lambda|{lc} G^2\rho^2 r^3\over\sigma^3}\;
{\td{r}\over r}
\label{ndotl}
\end{equation}
\begin{equation}
\ndotg = 4\pi \int\limits_\rcrit^\infty\, {G\M q \rho\over m_* \sigma}\; 
{\td{r}\over r}.
\label{ndotg}
\end{equation}

Before moving on, let us examine the various contributions to $\ndot$
in some more detail.  Let us define $f(r)$ to be the $r$-dependence of
the integrand in $\ndotl$, so
\begin{equation}
f(r) = {G^2\rho^2 r^3\over\sigma^3}
\label{fr}
\end{equation}
Now at small $r$, $\sigma\sim r^{-1/2}$, so $f(r)\to0$ as $r\to0$
provided $p<9/4$ (which is true in all the galaxies we consider).  The
lower limit in the first integral in principle should be finite, but
for $p<9/4$ we may set it to zero.  At large radius,
\begin{eqnarray}
\label{sigr1}
\sigma^2\sim r^{2-p}&, & p<3\\
\sigma^2\sim r^{-1/2}&, & p>3
\label{sigr2}
\end{eqnarray}
so $f(r)\to0$ as $r\to\infty$ provided $p>0$.  Thus generally $f(r)$
will reach a maximum at some radius $r|{max}$.  Quite generally we
expect that $r|{max}$ will be approximately equal to $r|a$ (for
$r|a<r|b$ or $0<p<9/4$ at large $r$) or $r|b$ (otherwise).  We can
identify two distinct regimes according to whether $\rcrit\grole
r|{max}$ .
\begin{enumerate}
\item If $\rcrit\gg r|{max}$ then $\ndotg$ will be insignificant
compared to $\ndotl$.\label{c1}
\item If $\rcrit\la r|{max}$ then $\ndotg$ will be at most comparable
with $\ndotl$.\label{c2}
\end{enumerate}
In case \ref{c1}
\begin{equation}
\ndot \approx \ndot|{max} = { N(r|{max})\over t|r(r|{max}) }
\label{ndotrmax}
\end{equation}
and in case \ref{c2}
\begin{equation}
\ndot \approx \ndot|{crit} = { N(\rcrit)\over t|r(\rcrit) }.
\label{ndotrcrit}
\end{equation}
Equations (\ref{ndotrmax}) and (\ref{ndotrcrit}) constitute a poor man's
recipe for calculating capture rates, and within a factor of order
unity are interchangeable with the full integral version in equation
(\ref{ndot}).

\subsection{Red giant capture}\label{rgsec}
For black hole masses $\ga2 \times 10^8\Msun$, the Schwarzschild
radius exceeds the tidal disruption radius for main sequence stars.
Red giants, as a result of their low densities, have much much larger
tidal radii, and can therefore be disrupted.  Red giants may enter the
loss cone in the same manner as main sequence stars: they diffuse
towards an empty loss cone for orbits below a critical radius, and
above that radius, they scatter onto a full loss cone. There is an
additional contribution to the red giant capture rate. Red giants
also grow onto the loss cone as they expand.  The loss cone for red
giants is larger than for main sequence stars ($q$ is larger because
they are less dense), so even when the main sequence loss cone is
empty, red giants are being born inside their own loss cone.  While
the total fraction of stars which are red giants at any given time may
be very small, the number of red giants susceptible to capture may be
significant.  We discuss each of these capture channels below. In
addition, throughout this discussion we use the term red giant loosely to
describe stars on both the RGB and AGB.

\subsubsection{Radius-time relationship \label{rtrel}}
In order to calculate the capture rates, it is necessary to understand
how red giants evolve in radius as a function of time because the loss
cone, $\theta|{lc} \propto R_\star$ (\ref{thlc}).  A good estimate of
the radius evolution can be easily determined as the radius and
luminosity of a red giant are largely independent of the stellar mass,
and depend only on the core mass.  Functional forms for $L(m_{\rm c})$
and $R(m_{\rm c})$ are given in Joss, Rappaport \& Lewis (1987) ,
where $m_{\rm c}$ is the core mass, which are good approximations for
both the red giant and ascending red giant branches.  Because the
dominant energy source in red giants is the $p{-}p$~chain with $\sim
0.7$\% efficiency and the core mass is increased as hydrogen burns to
helium, we may write
\begin{eqnarray}
L(m_{\rm c}) &\approx& 0.007 \Msun c^2 \; \pd{\mu}{t} \\ 
&\approx&  {10^{5.3}\mu^6\over 1+10^{0.4}\mu^4+10^{0.5}\mu^5} \Lsun,
\label{lumeq}
\end{eqnarray}
where $\mu=m_{\rm c}/\Msun$, and the right hand side comes from Joss,
Rappaport \& Lewis (1987).  Solving this equation for core mass as a
function of time, gives the luminosity-time dependence. A similar
expression to equation (\ref{lumeq}) yields the radius-time
dependence.

Applying this model between an initial core mass of 0.17~$\Msun$
and a final core mass of 0.8~$\Msun$ we find an initial radius of
3.09~$\Rsun$ and a maximum radius, $R_{\rm max}$, of 679~$\Rsun$ after
$7\times 10^8$~years. This maximum radius is in fact larger than the
radius predicted by stellar evolution codes, because mass loss 
significantly alters the evolution, especially when in the star
reaches the thermally pulsing AGB phase (TPAGB). We therefore limit the
red giant evolution at $200 \Rsun$ which corresponds to the onset of
the TPAGB (\eg{.} Bressan~\etal{.} 1992).

\begin{figure}
\epsfxsize=.9\hsize
\computelabs
\centerline{\epsfnormal{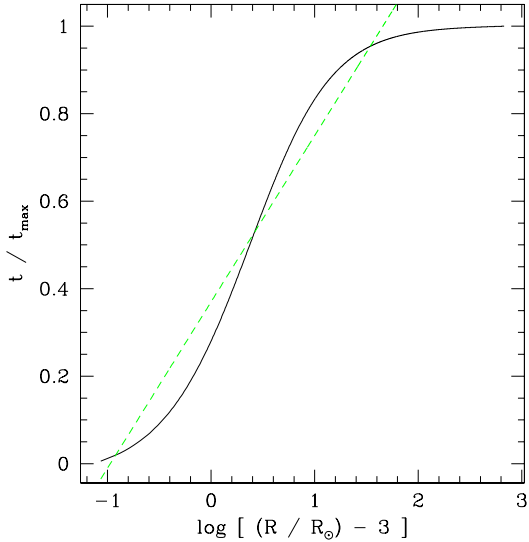}}
\vskip\baselineskip
\caption{ Time as a function red giant radius (solid line). The
dashed line shows an approximation adopted in eq.~\ref{tvsrapprox}.
}
\label{t-vs-r.ps}
\end{figure}

For our level of approximation of the red giant
capture rates, a simple analytical expression for the radius-time
dependence is sufficient. A course approximation
(figure~\ref{t-vs-r.ps}) is:
\begin{equation}
{t\over 7\times 10^8 {\rm years}} \approx 0.38
\ln\left( {R\over \Rsun}-3\right) +.37.
\label{tvsrapprox}
\end{equation}

\subsubsection{Red giant birth rate and number density}
In order to calculate the red giant capture rates, we must
determine their birth rate and their number density.

The main sequence lifetime of a star of mass $m$ is given
approximately by
\begin{equation}
t|{ms} \sim 10^{10} \lr{m\over\Msun}^{-2.5} \y
\label{tms}
\end{equation}
and the total number of stars, assuming a Salpeter mass function, is
\begin{equation}
\d{N}{m} \approx 1.6 \lr{m\over\Msun}^{-1.35},
\label{salpeter}
\end{equation}
where the total mass has been normalised to unity in the range
$(0.3,0.8)\Msun$.  Therefore the birth rate of red giants (per star)
is
\begin{eqnarray}
\d{\dot{N}|{new}}{N} & =& \d{N}{m}\;\mod{\d{m}{t}}\nonumber\\
& \approx & 0.65\times10^{-10} \lr{m\over\Msun}^{2.15} \y^{-1},
\label{rgbirth}
\end{eqnarray}
where $m$ is now the main sequence turn off mass (the mass of stars
which are currently being transformed into red giants).  The total
red giant lifetime $t_\rg\sim 7 \times 10^8\y$, so the fraction of
stars that are red giants (with radii between $\sim 3$ and $200\Rsun$)
is 
\begin{equation}
f_\rg = \d{\dot{N}|{new}}{N} \, t_\rg \approx 0.05.
\label{frg}
\end{equation}
We therefore expect that the red giant capture rate
will be lower than the main sequence rate by no more than a factor of
20. Another important quantity is the time averaged radius,
${<}R{>}\sim 12 \Rsun$. The evolution above $100\Rsun$ makes only a
small ($<$10\%) contribution to the time averaged radius.

\subsubsection{Maximum radius limits}
The red giant capture rates depend on the maximum radius reached by
the red giant. We consider how this radius may be truncated by stellar
collisions, and how this radius can effectively be truncated when the
red giant evolution time becomes shorter than the dynamical time.

First, consider the frequency of collisions between a red giant
and a main sequence star:
\begin{equation}
{t_{\rm coll}\over t_{\rm r}} \approx {3.7 \ln{0.4 N} \theta^2 \over
1 + 2 \theta},
\end{equation}
where $\theta$ is the Safronov~number, $G m_*/2 \sigma^2 R$ (see Binney
\& Tremaine 1987, equation 126).  for $\sigma \sim 100 \kms$, $\theta$
varies between about 5 and 0.01 as a red giant evolves. The collisions
are important if a red giant typically collides during its lifetime:
\begin{equation}
\int_{R|{min}}^{R|{max}} {{\rm d}t \over t|{coll} R}  \approx 1,
\end{equation}
where $R|{min}\approx 3\Rsun$.  Using the relations from
section~\ref{rtrel} and setting $\ln(0.4 N) \sim 20$, we find that
collisions are important when
\begin{equation}
t|r \la 1.1 \times 10^7 \left({ 250 \kms
\over \sigma} \right)^3 {\rm years}.
\end{equation}
For the systems we consider here, the shortest relaxation times
(at $r|a$) are many tens of times too large for collisions
to be important.

\begin{figure}
\epsfxsize=.9\hsize
\centerline{\epsfnormal{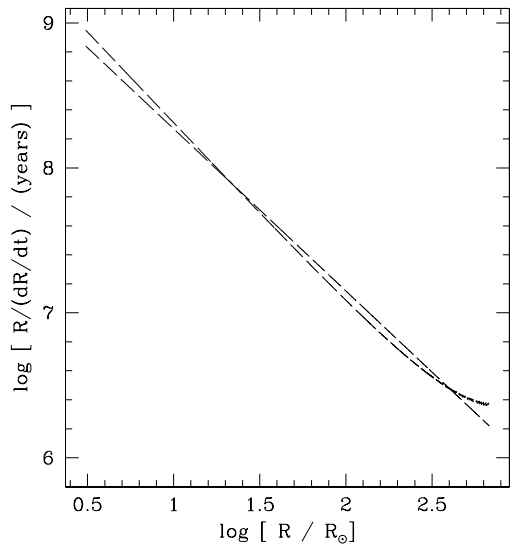}}
\vskip\baselineskip
\caption{Characteristic evolution time, $R/\dot{R}$ as a function
radius for red giants (solid line). The dashed line shows an simple
fit to the model.}
\label{rdotr}
\end{figure}

Second, at large system radii, the most extended red giants may evolve
on a time faster than the dynamical time, so that extended red giants
have a reduced chance of encountering the black hole.  Using the red
giant evolution model from Section \ref{rtrel}, we calculate the
characteristic evolution time, $R/\dot{R}$ as a function of radius
(Figure \ref{rdotr}).  We find that over the radii range
$3{-}200\Rsun$, the relationship is well approximated by
\begin{equation}
t_R\equiv R/\dot{R} \sim 6.5 \times 10^6 \left({R \over 200
\Rsun}\right)^{-1.12}.
\end{equation}
The maximum radius for a red giant is therefore the smaller of
$200\Rsun$ and the radius at which $t_R=t|d$.  In practice for the
Nuker galaxies in the regions of interest $t_R$ is always much greater
than $t|d$, and we take $R|{max}=200\Rsun$.

\subsubsection{Capture analogous to main sequence capture}
Main sequence capture occurs by diffusion onto an empty loss cone
at radii below the critical radius, and by scattering onto a full loss
cone above the critical radius (equations \ref{diffuse}, \ref{full}).
In general, the larger the radius of the star, the larger the critical
radius, and at large radii, the critical radius scales as the stellar
radius (equations \ref{rcrit}, \ref{sigr1}).  For red giants we must
define three regions because the red giant radius and therefore the
critical radius are functions of time.  Below a radius $\raa$
determined by solving equation (\ref{rcrit}) with $q = 3 q|{ms}$, the
loss cone is depleted on a dynamical time for all red giants, so the
diffusion rate is appropriate.  Beyond the second radius $\r2c$
determined by solving equation (\ref{rcrit}) with $q = 200 q|{ms}$,
the loss cone is full, so the full loss cone rate is appropriate.
Between these two radii, the full rate is appropriate until the star
reaches a size at which it can scatter entirely into or out of the
loss cone in one dynamical time, \ie{.} the stellar radius $R_\tr$ for
which equation (\ref{rcrit}) is satisfied.  The rate of captures of
red giants from these three contributions can therefore be written:
\begin{eqnarray}
\d{\ndot_\rg}{N} = &&\nonumber\\ 
&&\kern-45pt \d{\dot{N}|{new}}{N}
\left[ \int_{R|{min}}^{R_\tr} {R\over \Rsun} {\td{t}\,\thlc^2 \over t|d} +
 \int_{R_\tr}^{R_{\rm max}} {\td{t}\over \ln(2/\thlc) t|r}
 \right].
\label{dndotrgms}
\end{eqnarray}
The limits in the integrals are understood to mean the time
at which $R$ is equal to the value specified.  The critical red-giant
radius $R_\tr$ is equal to $R|{min}$ for $r<\raa$ and to $R|{max}$ for
$r>\r2c$.

\subsubsection{Captures from growth onto the loss cone\label{growth}}
There will be some maximum radius for red giants $R|{max}$, fixed
either by stellar evolution (at $\sim 200 \Rsun$), or by other means
\eg{,} the rate at which their envelopes are stripped by collisions
with other stars.  As we discuss below, the appropriate maximum radius
is almost always set by stellar evolution.  This maximum radius
defines a loss cone with angular size
\begin{equation}
\theta_\rg^2 = \thlc^2 {R|{max}\over \Rsun},
\label{thrg}
\end{equation}
where (as before) $\thlc$ is the angular size of the main sequence
loss cone.  Assuming that all stars which become red giants inside
this loss cone will be captured, the rate of red giant
captures from a given radius (per star) is
\begin{equation}
\d{\ndot_\rg}{N} = {\theta_\rg^2 \d{\dot{N}|{new}}{N}} \approx
{460\thlc^2\over t_0} \approx
{4.6\times10^{-8} \thlc^2}\y^{-1},
\label{dndotrg}
\end{equation}
where $t_0$ is the age of the galaxy, and we have chosen
$R|{max}=200\Rsun$. 

\subsubsection{Total capture rate \label{totalcapture}}
The rate of red giant capture per star is the sum of equations
(\ref{dndotrgms}) and (\ref{dndotrg}):
\begin{eqnarray}
\d{\ndot_\rg}{N} &=&
\dts{\dot{N}|{new}}{N}
  \left[ \int_{R|{min}}^{R_\tr} {R\over \Rsun} {\td{t}\,\thlc^2 \over t|d} + \right. \nonumber\\ 
&& \;\;\left. \int_{R_\tr}^{R_{\rm max}} {\td{t}\over \ln(2/\thlc) t|r}
 \, + \, {R|{max}\over\Rsun} \thlc^2 \right],
\label{dnotrgtot}
\end{eqnarray}
which must be integrated over the system to obtain the total capture
rate
\begin{equation}
\ndot_\rg = 4\pi \int_0^\infty \d{\ndot_\rg}{N} \rho r^2\td{r}.
\label{ndotrg}
\end{equation}
For the Nuker galaxies, since $\rcrit$ is generally large for
main-sequence stars it is larger still for red giants, and
$R_\tr=R|{min}$.  We also find that growth onto the loss cone
(equation \ref{dndotrg}) dominates the red giant capture rate.

\subsection{Calibration}
\FR{} considered a system in which the stars inside the sphere of
influence have relaxed to the `zero-flow' distribution of Bahcall \&
Wolf (1976) with $p=7/4$, and outside $r|a$ the stars are
approximately isothermal with a homogeneous core of radius $r_0$ and
density $\rho_0$.  One case they concentrated on was that in which
$\rcrit<r|a<r_0$.  \CK{} also calculated the capture rate (as well as
the full anisotropic distribution function) in a system with
$\rcrit<r|a$ and a relaxed cusp with $p\approx1.8$.  They give a
scaling relation (their equation 66) for $\ndot$ in these
circumstances which is substantially the same as \FR{} equation (16a),
except for the normalisation.  We use this normalisation to calibrate
our model capture rates.


\begin{table*}
 \centering

  \caption{Observational data and summary of results. Column (2) taken
  from Byun \etal{.} 1996; Columns (3-4) from Magorrian \etal{.}
  1998.  Column (1), name of galaxy; (2), $r|b$ in \pc{}; (3), Nuker
  law indices $(\alpha,\beta,\gamma)$; (4) $\log_{10}\M$ black hole
  mass; (5) $r|a/r|b$; (6) $r|a$ in arcsec; (7) main-sequence capture
  time ($\ndot$/\y); (8) red giant capture time ($\ndot_\rg$/\y); (9)
  projected stellar luminosity in central arcsec $\ell(1) =
  L(r<1")/L|{Edd}$. }

  \begin{tabular}{@{}lrcrrcrcc@{}}
   NGC & $r|b$   &$(\alpha,\beta,\gamma)$
   & $\log\M$ & $r|a\over r|b$ & 
   $r|a(")$ & $-\log \ndot$ & $-\log \ndot_\rg$ & $-\log \ell(1)$ \\
1399 & $269.15$ & $(1.50, 1.68, 0.07)$ &
$ 9.72$& $ 0.38$& $ 1.19$& $ 6.13$& $ 4.23$ & $ 4.92$ \\
1600 & $1258.93$ & $(1.98, 1.50, 0.08)$ &
$10.06$& $ 0.09$& $ 0.48$& $ 6.60$& $ 3.92$ & $ 5.02$ \\
2832 & $398.11$ & $(1.84, 1.40, 0.02)$ &
$10.06$& $ 0.51$& $ 0.46$& $ 6.41$& $ 4.09$ & $ 4.36$ \\
3115 & $117.49$ & $(1.47, 1.43, 0.78)$ &
$ 8.55$& $ 0.07$& $ 0.19$& $ 4.80$& $ 4.46$ & $ 3.47$ \\
3377 & $ 4.37$ & $(1.92, 1.33, 0.29)$ &
$ 7.79$& $ 0.97$& $ 0.09$& $ 4.80$& $ 5.10$ & $ 2.99$ \\
3379 & $83.18$ & $(1.59, 1.43, 0.18)$ &
$ 8.60$& $ 0.22$& $ 0.38$& $ 5.76$& $ 5.00$ & $ 4.02$ \\
 608 & $27.54$ & $(1.05, 1.33, 0.00)$ &
$ 8.39$& $ 0.54$& $ 0.15$& $ 5.54$& $ 5.10$ & $ 3.51$ \\
 168 & $446.68$ & $(0.95, 1.50, 0.14)$ &
$ 9.08$& $ 0.19$& $ 0.48$& $ 6.47$& $ 5.00$ & $ 4.11$ \\
4467 & $239.88$ & $(7.52, 2.13, 0.98)$ &
$ 7.44$& $ 0.04$& $ 0.13$& $ 5.36$& $ 5.76$ & $ 3.13$ \\
4472 & $177.83$ & $(2.08, 1.17, 0.04)$ &
$ 9.42$& $ 0.17$& $ 0.41$& $ 6.05$& $ 4.20$ & $ 4.93$ \\
4486 & $562.34$ & $(2.82, 1.39, 0.25)$ &
$ 9.55$& $ 0.14$& $ 1.06$& $ 6.32$& $ 4.39$ & $ 5.21$ \\
4486 & $13.49$ & $(2.78, 1.33, 0.14)$ &
$ 8.96$& $ 9.95$& $ 1.81$& $ 5.90$& $ 5.02$ & $ 4.38$ \\
4552 & $47.86$ & $(1.48, 1.30, 0.00)$ &
$ 8.67$& $ 0.24$& $ 0.15$& $ 5.57$& $ 4.75$ & $ 3.80$ \\
4564 & $38.90$ & $(0.25, 1.90, 0.05)$ &
$ 8.40$& $ 0.37$& $ 0.20$& $ 5.24$& $ 4.90$ & $ 3.34$ \\
4621 & $218.78$ & $(0.19, 1.71, 0.50)$ &
$ 8.45$& $ 0.03$& $ 0.10$& $ 4.74$& $ 4.52$ & $ 3.09$ \\
4636 & $239.88$ & $(1.64, 1.33, 0.13)$ &
$ 8.36$& $ 0.07$& $ 0.24$& $ 6.24$& $ 5.35$ & $ 4.07$ \\
4649 & $263.03$ & $(2.00, 1.30, 0.15)$ &
$ 9.59$& $ 0.29$& $ 1.04$& $ 6.19$& $ 4.32$ & $ 5.11$ \\
4874 & $1202.26$ & $(2.33, 1.37, 0.13)$ &
$10.32$& $ 0.39$& $ 1.04$& $ 6.93$& $ 4.13$ & $ 5.12$ \\
4889 & $758.58$ & $(2.61, 1.35, 0.05)$ &
$10.43$& $ 0.64$& $ 1.07$& $ 6.72$& $ 3.96$ & $ 4.89$ \\
6166 & $1202.26$ & $(3.32, 0.99, 0.08)$ &
$10.45$& $ 0.68$& $ 1.50$& $ 7.16$& $ 4.18$ & $ 5.42$ \\
7332 & $75.86$ & $(4.25, 1.34, 0.90)$ &
$ 6.84$& $ 0.01$& $ 0.01$& $ 4.10$& $ 4.81$ & $ 1.31$ \\
7768 & $199.53$ & $(1.92, 1.21, 0.00)$ &
$ 9.96$& $ 1.25$& $ 0.50$& $ 6.34$& $ 4.17$ & $ 4.25$ \\
 221 & $ 1.47$ & $(2.00, 1.28, 0.53)$ &
$ 6.36$& $ 0.50$& $ 0.19$& $ 4.53$& $ 5.27$ & $ 3.05$ \\
 224 & $288.40$ & $(1.12, 1.52, 0.33)$ &
$ 7.79$& $ 0.04$& $ 2.80$& $ 5.95$& $ 5.76$ & $ 5.41$ \\
  MW & $ 0.38$ & $(1.80, 0.80, 0.00)$ &
$ 6.42$& $ 1.23$& $11.38$& $ 4.32$& $ 5.07$ & $ 6.80$ \\
\end{tabular}
\label{dattab}
\end{table*}

\begin{figure}
\epsfxsize=.9\hsize
\computelabs
\ylabel{\mu|V\,(\magsec)}{-.05}{-.5}
\xlabel{r\,(\hbox{arcsec})}{.5}{-.95}
\centerline{\epsfnormal{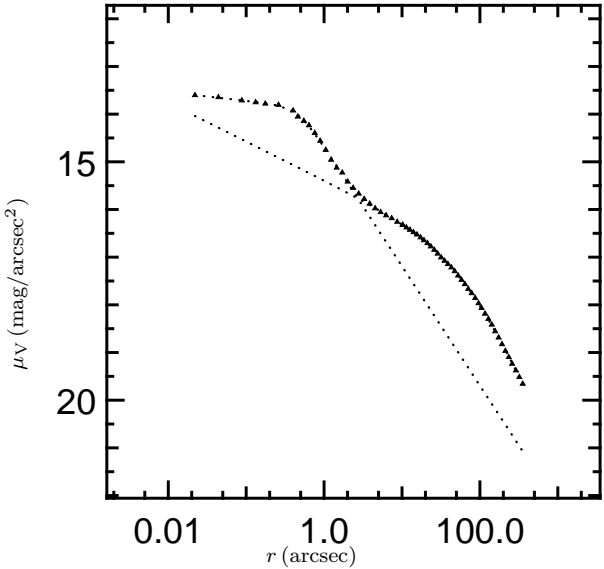}}
\vskip\baselineskip
\caption{ Deconvolved V-band surface density of M31 (triangles) with
Nuker law fits to the inner ($r<1.75"$) and outer ($r>3.6"$) profiles
(dotted lines).}
\label{m31fig}
\end{figure}

\section{The Nuker galaxies}\label{datsec}
To calculate a capture rate, we need to know the density profile
$\rho(r)$ and the black hole mass $\M$.  Given the observed surface
brightness $I(r)$, we can derive $\rho(r)$ by an Abel inversion,
assuming a constant mass-to-light ratio $\Upsilon$.  We do not assume
anything about the density profile in regions which are not resolved
by the observations, merely extrapolating the profiles inwards.  This
will not affect the capture rates provided min$[\rcrit,r|{max}]$ is
not much smaller than the resolution limit of the observations, which
turns out to be the case.  Most of the data we use are quoted as a
Nuker law fit to $I(r)$ (Byun \etal{.}  1996).  Thus $I(r)$ is in the
form of two power laws, separated by a break radius $r|b$, and
$\rho(r)$ has a similar form, with a break radius close to $r|b$.  The
data we have used are reproduced in Table \ref{dattab}, which also
lists some interesting derived quantities. M32 is listed as NGC\,221,
M31 is NGC\,224, and the Milky Way is `MW'.  The Milky Way data is
from the model by Genzel \etal{.} 1996.  The Nuker parameters for M32
are taken from Lauer \etal{.} (1992).  For M31 we fit the photometry
ourselves using data from Lauer \etal{.} (1993) ($r<10"$) and Kent
(1987) ($r>10"$) and matching at $r=10"$.  The resulting surface
density profile is shown in Figure \ref{m31fig}, which also shows the
Nuker law fits listed in Table \ref{dattab}.  We fit the nuclear
component ($r<1.75"$) separately, and name it NGC\,224N in Table
\ref{dattab}.  This component is not self gravitating and is probably
rotationally supported (Lauer \etal{.}  1993, Tremaine 1995), and the
capture rate is therefore almost certainly much less than in our
isotropic model.  We calculate the latter in order to satisfy ouselves
that the capture rate is not dominated by the nuclear component, which
is confirmed by the results.  The scale of the flattened component in
M31 is much smaller than the physical scale which dominates stellar
capture rates.  Similarly, in the other galaxies the capture rate is
dominated by a physical radius which is well resolved, so unresolved
substructure should not affect our results.

Throughout this section we assume where necessary that the mass of the
black hole is proportional to the mass of the galaxy (or spheroid)
with constant of proportionality $\approx .006$ (Magorrian \etal{.}
1998), and the mass to light ratio obeys the fundamental plane
relation $\Upsilon\propto L^{0.2}$.  Thus the black hole mass is
\begin{equation}
\M \approx 6\times10^8 \lr{L\over L_*}^{1.2}.
\label{mvl}
\end{equation}

\begin{figure}
\epsfxsize=.9\hsize
\computelabs
\ylabel{\ndot (\y^{-1})}{-.01}{-.5}
\xlabel{\M (\Msun)}{.5}{-.95}
\centerline{\epsfnormal{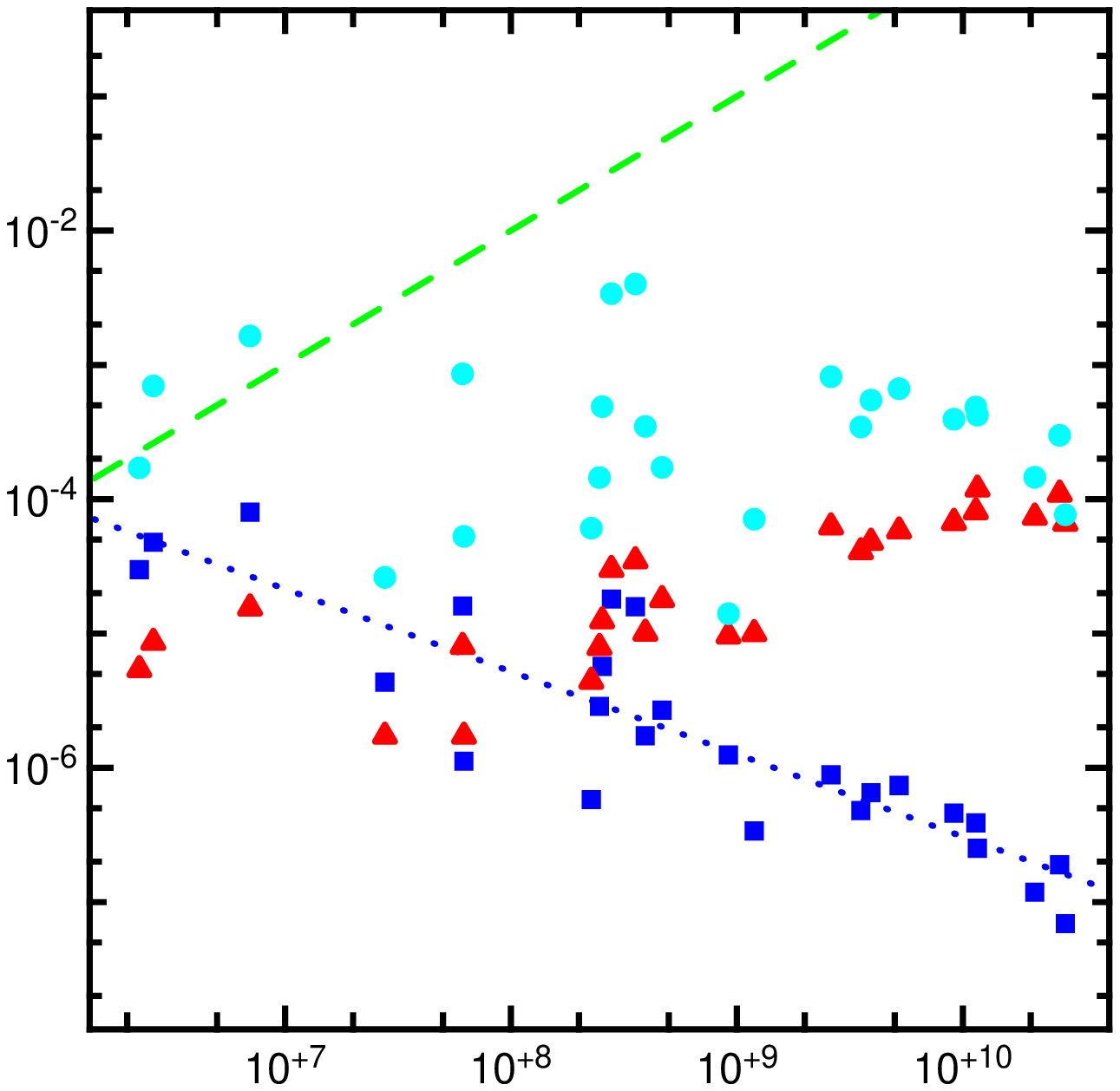}}
\vskip\baselineskip
\caption{ The capture rates $\ndot (\y^{-1})$ for main sequence stars
(squares) and red giants (triangles) as a function of black hole mass.
The circles are the main sequence capture rates in the case where loss
cones are all full.  The dotted line is the best fit to the main
sequence rates (with loss cone), and has slope $-0.61$.  The dashed
line is the upper limit set by $\ndot m_* = \M/t_0$ with $t_0=10^{10}\y$.  }
\label{ndotm}
\end{figure}

\subsection{Capture rates}\label{detsec}
Figure \ref{ndotm} shows the capture rates as a function of the black
hole mass.  In our models all the Nuker galaxies have $\rcrit\gg r|b$
and all but two have $r|b\ga r|a$, and hence $r|{max}\approx r|a$.
This accounts for why both the main sequence rate and the red giant
rate are correlated with $\M$.  In the red giant case, the capture
rate is proportional to the mass within $r|a$ which is of order the
black hole mass.  In the main sequence case the fuelling rate is
proportional to this mass divided by $t|r(r|a)$.  The fundamental
plane leads to a correlation between $t|r(r|a)$ and $M$, and hence to
the correlation in Figure \ref{ndotm}.  We can also calculate
$\ndot|{max}$ equation (\ref{ndotrmax}), and the result is well
correlated with the value of $\ndot$ given in Table \ref{dattab} (over
a range in $\ndot$ of 3 orders of magnitude $\ndot|{max}$ is within a
factor of two or so of $\ndot$).  We can also use equation
(\ref{ndotrmax}) to estimate what the effect on $\ndot$ would be of
errors in the determination of the black hole mass.  The result for
main sequence stars and empty loss cones is that errors in
$\M$ tend to move $\ndot$ roughly {\em along} the $\ndot{-}\M$
relation in Figure \ref{ndotm} (dotted line).

\subsubsection{Comparison with supernova rate}
Main sequence capture is faster for smaller black holes, and this
leads us to expect that the total rate of capture in the local
universe will be dominated by the galaxies with smallest black holes.
To see this more clearly, let us compare the rate of star capture with
the supernova rate in the local universe.  We assume a galaxy
population with luminosity function of the form
\begin{equation}
\d{N}{L} \propto {1\over L} \lr{L\over \Lstar}^\alpha \exp\lr{-L\over\Lstar},
\label{schecter}
\end{equation}
(Schechter 1976) where empirically $\alpha$ is typically small and
negative.  We adopt $\alpha=-0.07$, and $\Lstar=1.8\times10^{10}\Lsun$
(Estafthiou \etal{.} 1988).  The supernova rate in a galaxy is
roughly proportional to its luminosity $L$.  In the local universe it is
approximately
\begin{equation}
\dot{N}|{SN} \approx 10^{-2} {L\over \Lstar} \y^{-1},
\label{snrate}
\end{equation}
and the rate of capture of stars from Figure \ref{ndotm} and equation
(\ref{mvl}) is
\begin{equation}
\ndot \approx 2\times10^{-6} \lr{L\over \Lstar}^{-s} \y^{-1}.
\label{ndotfit}
\end{equation}
Together with the fundamental plane relation between $\Upsilon$ and
$L$, Figure \ref{ndotm} gives $s\approx 0.61\times1.2 = 0.73$.  The
ratio of the star capture rate to the supernova rate is
\begin{equation}
f_* \approx { \displaystyle \int \ndot(L) \d{N}{L} \td{L} \over
\displaystyle \int \dot{N}|{SN}(L) \d{N}{L} \td{L} },
\label{fstar}
\end{equation}
whence 
\begin{equation}
f_* \approx 2\times10^{-4}\;
{\alpha+1\over s-\alpha} \lr{\Lstar\over L|{min}}^{s-\alpha}
\label{fstarn}
\end{equation}
where $L|{min}$ is the smallest luminosity for which we think the
majority of galaxies have a massive black hole.  

The value of $L|{min}$ is not determined by observations, since most
of the observations are limited to larger systems.  M32
($L=3.7\times10^{8}\Lsun$) is the smallest galaxy in the Nuker sample
with a confirmed black hole, but it is also an unusually dense system
at this luminosity.
Dwarf galaxies in general are usually thought not
to contain black holes.
For the sake of argument consider
$L|{min}=10^8\Lsun$, from which equation \ref{fstarn} predicts that
for every supernova there would be $\sim0.01$ tidal disruptions.  If
we take $L|{min}=10^7\Lsun$ that number goes up to $\sim 0.1$.  With
searches now detecting many tens of supernovae, these numbers suggest
that the same surveys or similar ones at shorter wavelength should
uncover tidal disruptions at an interesting rate.

The rate of red giant captures $\ndot_\rg$ is approximately
proportional to $M$.  {}From Figure \ref{ndotm} we can read off the
constant of proportionality and use equation (\ref{mvl}) to write
\begin{equation}
\ndot_\rg \approx 10^{-5} \lr{L\over \Lstar}^{t} \y^{-1}.
\label{ndotrgfit}
\end{equation}
with $t\approx 1.2$.  However, some of the corresponding tidal
disruption events will have a very long timescale, and hence may not
be detected as flares.  Suppose we can
only observe flares in cases where the return time for tidal debris
$t|{min}$ is less than 10\y ~(see Ulmer 1997 for details).  Then the
maximum radius of a red giant whose disruption would be observable is
\begin{equation}
R|{vis} \approx 100 \Rsun \lr{t|{min}\over10\y}^{2/3} 
\lr{\M\over 10^6\Msun}^{-1/3}\;.
\label{rgrmax}
\end{equation}
Since red giant disruption is dominated by stars growing onto the loss
cone , the rate of disruptions up to $R|{vis}$ is proportional to
$R|{vis}$, and thus
\begin{equation}
\ndot|{vis} \approx 6\times10^{-7} \lr{t|{min}\over10\y}^{2/3} 
\lr{L\over \Lstar}^{2t/3} \y^{-1}.
\label{ndotvis}
\end{equation}
Integrating over the luminosity function we obtain
\begin{equation}
f|{*vis} \approx 6\times10^{-5} \lr{t|{min}\over10\y}^{2/3} 
	\;{\alpha+1\over 2t+3\alpha},
\label{fstarrg}
\end{equation}
where $f|{*vis}$ is the ratio of the rate of visible red giant disruptions to
the supernova rate, 
so the red giant capture rate is probably much less than the main
sequence capture rate in the local Universe.
Note however that it is not
possible to rule out intermittent behaviour in the accretion of tidal
debris (e.g. Lee, Kang, \& Ryu 1996)
and so equation (\ref{fstarrg}) represents a lower limit to
the rate of detectable red giant disruptions.

\subsubsection{Upper limits to capture rate}
An upper limit to the capture rate is often given as the full-loss
cone rate in the region $r<\rcrit$.  If non-axisymmetric processes can
keep the loss-cone full the capture rate can be approximately derived
from equation (\ref{full}).  The rate of capture of stars from the
full loss cone would be given by $\ndotg$ but integrated all the way
down to $r=r|a$ (assuming that the black hole itself imposes enough
symmetry to keep the loss cone starved at $r<r|a$).  The dominant
radial scale is thus that at which $\rho/\sigma$ is largest.  Since
$\sigma\sim r^{-1/2}$ for small $r$, we see that if $\rho\sim r^{-p}$
then $\ndot$ is dominated by the contribution from $r|a$ if $p>1/2$.
This is the case in our models in all but three cases (and a more
realistic model of the bound stars inside $r|a$ would probably also
have $p>1/2$).  The full loss cone rates for the Nuker galaxies are of
order $10^{-3}\y^{-1}$, roughly independent of $\M$ (circles in Figure
\ref{ndotm}).  Thus the capture rate
would be roughly
\begin{equation}
\ndot m_*={\M/t_0} \approx 4\times10^{-4} \lr{L\over \Lstar}^{s} \Msun\y^{-1},
\label{ndotmq}
\end{equation}
with $s\approx0.2$, which is larger than the supernova rate for $L$
less than a few percent of $L_*$.  Substituting equation
(\ref{ndotmq}) in equation (\ref{fstar}), we find
\begin{equation}
f_* \approx  4\times10^{-2}\;{\alpha+1\over s-\alpha} \lr{L|{max}\over
\Lstar}^{s-\alpha}
\label{fstarmq}
\end{equation}
where $L|{max}$ is the largest luminosity galaxy which can tidally
disrupt a star (as opposed to swallowing it whole).  Taking the
maximum black hole mass for tidal disruption to be $10^8\Msun$ we
obtain $f_*\sim 0.1\,$.  This means that tidal disruptions would be about
ten times less likely than supernovae if the loss cone was full.  The
tidal disruptions in this case would also be coming mainly from
galaxies with $L\sim L|{max}$.  The red giant capture rate would be
simply $f_\rg R|{max}/\Rsun\approx10$ times the main sequence rate
(with $L|{max}\approx L_*$).  

Equations (\ref{fstarmq}) and (\ref{fstarn}) together provide a
constraint on the mass supply for the black holes.  If only a small
number of tidal disruptions were observed, and they came from the
faintest galaxies, then the dominant fuelling would be via empty loss
cones.

Another upper limit to the capture rate is given by assuming that the
galaxies we see today are not in a special state, so their capture
rates must be smaller than $\M/t_0$ (the dashed line in Figure
\ref{ndotm}).  In fact this is a natural value for the capture rate in
the case that the black holes grew by eating stars from the cluster
around them.  For the majority of the Nuker galaxies $\M/t_0$ is
actually {\em larger} than the full loss cone rate.  Thus there has
not been enough time for the black holes to grow by eating stars from
the nuclear star cluster.  A more efficient mechanism must be
responsible for their having grown to the present masses.  Merrit \&
Quinlan (1997) have argued that the universal $x\approx 0.006$ is a
result of the black hole eating stars until the intrinsic triaxiality
of the star cluster is reduced by the presence of the black hole
(Gerhard \& Binney 1985). At least for the larger galaxies,
triaxiality could never supply enough stars to provide all the mass in
the black holes.

\begin{figure}
\epsfxsize=.9\hsize
\computelabs
\ylabel{r|a (")}{.0}{-.5}
\xlabel{\M (\Msun)}{.5}{-.95}
\centerline{\epsfnormal{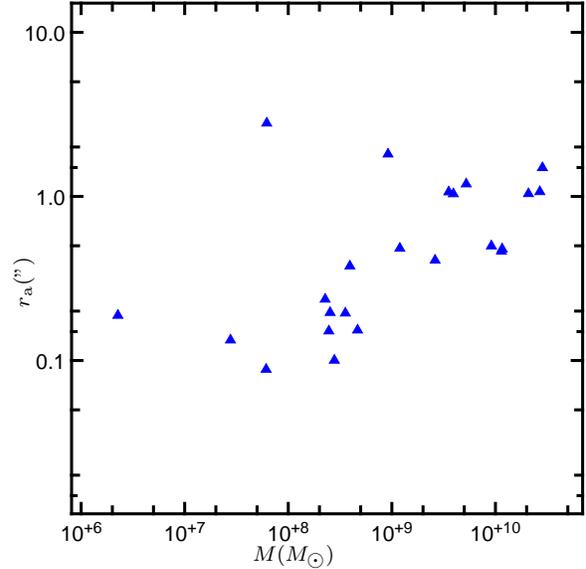}}
\vskip\baselineskip
\caption{ The radius of the sphere of influence $r|a$ in arcsec as a
function of black hole mass.
}
\label{rafig}
\end{figure}

\subsection{Keplerian regime}
The criticism that $r|a$ is often comparable with the resolution limit
of the observations in question (Rix 1993) is no longer valid.  This
has been pointed out by Kormendy (1994), and we support his
view with our carefully calculated values of $r|a$ in the sense that
the measured values do not cluster around any particular value (Table
\ref{dattab} and Figure \ref{rafig}).  On the other hand, some of the
smallest black holes have $r|a$ comparable to or smaller than $0.1"$.

The black hole masses measured by Magorrian \etal{.} (1998) required
detailed modelling of the kinematics of the surrounding galactic
nuclei.  In principle one should see at small radius the Keplerian
regime in which $\sigma^2\sim \M/r$.  This would be a very clear
signature of a black hole, giving directly the mass to within a factor
of order unity.  The Keplerian regime should occur within some radius
of order $r|a$.  We find that the logarithmic slope of $\sigma^2$
reaches $-0.95$ at a radius $r|K$ which is generally much less than
$r|a$, with median value $\approx 0.2 r|a$ (this is of course
consistent with the fact so much modelling had to be done to measure
the black hole masses).  This situation is also seen in the centre of
the Milky Way, where the Keplerian regime is now resolved
(Genzel \etal{.} 1996) at $r|K\sim r|a/10$.

One might expect that the galaxies with the largest values of $r|K$
would have the most secure black hole detections, but this appears not to
be the case.  The three galaxies where we find that $r|K>0.5"$ are
NGC\,4486B, NGC\,7768 and M31, and these do not appear to have
significantly better black hole masses than any of the others.

\section{Discussion}\label{discuss}
Note that the scaling relations of \FR{} and \CK{} giving $\ndot$ in
terms $\M$, $\rho_0$ and $r_0$ apply only in very specific
circumstances.  The result of \CK{} only applies to the case $\rcrit<
r|a$.  If $\rcrit<r|a$ {\em and} $p\approx1.8$ for $r<r|a$ then \CK{}
equation (66) can be used to calculate $\ndot$, replacing
$\rho_0\to\rho(r|a)$ and $r_0\to r|a$, with the result 
roughly independent of what happens at $r>r|a$.  \FR{} are careful to
distinguish between the cases $\rcrit\grole r|a$, and in these cases
they use $p=7/4$ or $p=0$ accordingly.  It is straightforward to
generalise their expressions using equations (\ref{ndotrcrit}) and
(\ref{ndotrmax}).

Rauch \& Tremaine (1997) have discussed the enhancement of tidal
disruption rates by a process they call `resonant relaxation'.  This
effect can increase tidal disruption rates of stars which are {\em
bound } to the black hole (\ie{.} at $r<r|a$) by factors of order
unity.  Detailed calculations by Rauch \& Ingalls (1997) also show
that the effect is also quencehed by relativistic precession for black
hole masses $\M\ga10^8\Msun$.  In the models we constructed the
fuelling rate was never dominated by strongly bound stars, so in
common with Rauch \& Ingalls (1997), we find that resonant
relaxation has little effect on tidal disruption rates.

There have been some observations of active galaxies which develop a
broad-lined feature in their spectrum over a period of the order of
months (Storchi-Bergmann \etal{.} 1995, Eracleous \etal{.} 1995).
Eracleous \etal{.} (1995) constructed a phenomonological model of such
an event which involved an elliptical accretion disc, and suggested
that such a disk could arise from the tidal disruption of a star (see
also Syer \& Clarke 1992).  The size of the disc they required to fit
the observations was too large to correspond to a main sequence
disruption. It could, however, arise from the disruption of a red
giant.  Our discussion in Section \ref{detsec} concluded that red
giant disruptions with such short timescales should be very rare, so
it is perhaps not surprising that such events are not encountered
frequently. 

\section*{Acknowledgments}
We thank Martin Rees and Achim Wei\ss{} for helpful discussions and
comments.

{}

\bsp
\label{lastpage}
\end{document}